\documentstyle{article}
\begin{document}

\bigskip
\centerline{\bf EXOTIC ATOM FORMATION IN EXTERNAL ELECTRIC FIELDS}
\bigskip
\centerline{Jaroslaw Kulpa , Slawomir Wycech }
\centerline{Soltan Institute for Nuclear Studies }
\centerline{  Hoza 69, 00-681 Warsaw,Poland}
\bigskip
\centerline{\bf Abstract}

 Validity of the coalescence model of exotic atom
formation in the ion collisions is studied . An estimate is given
for the rates of P state production in the electric fields of the
colliding ions . Possibilities of quantum beat effects are indicated .

\section{Introduction}          % THIS COMMAND MAKES A SECTION TITLE.

    We study possibilities of production of  pairs composed of an
elementary particle and its antiparticle in the atomic bound
states.The particles may be produced in proton or ion collisions and the
simplest cases of practical interest are the lightest pairs
$e^{+} \ e^{-}$ (positonium), $\pi^{+} \ \pi^{-}$ (pionium) or
$K^{+} \ K^{-}$ (kaonium). These atoms are large systems when compared
to the  regions where the pairs are created . Thus , amplitudes for the
formation of atoms in final states are  proportional to the atomic
wave functions at the origin and the pair creation amplitudes . This
approximation known as the coalescence model yields  rates of mesic
atom formation high enough to allow experiments . Thus the search
for pionium has been undertaken or contemplated at CELSIUS,CERN,
COSY,Indiana UCL, [1,2,3,4,5].

     The main physical motivation  to study pionium (or kaonium)
is a measurement of its lifetime . In the S states the latter
may be expressed by the appropriate  particle-antiparticle
scattering length [6] , a quantity of great theoretical interest .
The main difficulty of these experiments
is that the decay of the 1S state , due to strong interactions , is very
fast ( about $10^{-13}$ sec for pionium). In order to find the lifetime
one has to compare the main $\pi^{0}\ \pi^{0}$ decay mode with very
rare two gamma decay mode .
In principle ,the atom's life may be prolonged by orders of magnitude
if it is formed in the P states that decay slowly to the ground state
by the X-ray emission [7] . This would open new experimental
possibilities and a new scale given by the calculable X-ray decay rate .
However, a direct P wave pair production with the subsequent P
state atom formation in a coalescent manner is very unlikely [6] .
In this paper another mechanism for the P wave atom production
is considered . We assume the pair to be produced
at short ranges in an S wave . Subsequent formation of the P state atoms
is achieved by a Stark mixing in the external electric field of the colliding
particles . A related problem discussed here is the rate
of destruction of the S state atoms formed via the coalescence mechanism
and destroyed by the external electric field . The second question is
relevant to the studies of atomic production in the proton-nucleus
collisions at high energies [4] .

    The atom formation is discussed here as a problem of a pair created
in an external  static Coulomb field . In the full complication it
becomes a formidable problem of three body  Coulomb interactions in the
continuum.Only part of it has been solved with  satisfactory precision.
In particular we consider only atoms moving with high velocities.
In a kind of quasi-clasical solution we find the rate of destruction
of the 1S atoms in the external field and the rate of production of the
P wave states. The former is significant the latter is rather small.
An upper limit for these quantities is given  with a help of completeness
approximation . A system of succesive approximations is shown to
converge to this limit.

   This model shows that the participation of different atomic states
depends strongly on the final momentum of created pairs. If only
a limited number of final states is allowed the atomic system displays
effects of quantum beat . The atomic pair in the external, polarising,
electric field forms quasi-stationary Stark type combinations. As a
consequence the  probability of production of each atomic state
oscillates with respect to the final atom momentum. This phenomenon
is rather unstable against the approximations.
We believe ,it deserves more extended study as it may be related to the
production of correlated   $e^{+} \ e^{-}$  pairs seen at GSI ,[8] ,
in the heavy ion collisions.

\section{The model of atomic formation}    % THIS COMMAND MAKES A SUBSECTION
%%TITLE.

   This section presents  basic assumptions of this model. The equations
that describe the atom formation are derived and  simpified solutions are
found. The latter allow to form an intuitive picture of the formation
mechanism. An approximation scheme for the high energy region is derived.

    In  cases of practical interest the atomic Bohr radii  , B ,
are much larger than radii of the interaction region where the pair
is created. This leads to the expectation that  the atom formation amplitude
A(E) is proportional to the wave function at the origin     :

\begin{equation}
A(E)= F\psi(\vec{\varrho}=0,\vec{r}=0)
\end{equation}
where F is an amplitude for the pair creation .

    This paper is devoted entirely to the determination of the final
state wave function ${\psi}$. The latter depends on two coordinates :
$\vec{\varrho}$ that describes the center of mass of the pair
$\pi^{+}\ \pi^{-}$, and $\vec{r}$ that is  the relative coordinate
in the center of mass. We assume the adiabatic situation i.e. fixed ions
and fast mesons . Then the external polarization potential V of colliding
nuclei acting on the pair becomes :

\begin{equation}
V=Z \alpha(\frac{1}{\mid\vec{\varrho}-\frac{\vec{r}}{2} \mid}-\frac{1}
{\mid\vec{\varrho}+\frac{\vec{r}}{2} \mid})
\end{equation}

where Z is the total charge of the nuclei.

    Let us  start with the  Lippman-Schwinger integral equation describing
nonrelativistic dynamical formation of atomic states of the pair
$\pi^{+} \pi^{-}$. We have,

\begin{equation}
    \psi = \psi_{0} +\int G V \psi d^{3}\varrho d^{3}r
\end{equation}
and the boundary conditions , corresponding to the ingoing waves ,
are to be specified later .In this equation the "internal " interactions
within the pair have been solved and described by the Green's
function G . At first only the discrete terms of this Green's function are
taken into account.

\begin{equation}
    G \equiv G(\vec{r},\vec{\varrho},\vec{r'},\vec{\varrho'}) \approx
    \sum_{nlm} \mid\varphi_{nlm}>< \varphi_{nlm}\mid G(\vec{\varrho},
    \vec{\varrho'},E-E_{n}),
\end{equation}
where  E is the total energy of the pair . We are interested in the high
kinetic energies  ( MeV region )  and  neglect the atomic
binding energies  \( E_{n}\) ( KeV region ).  Then ,
\(  E=\frac{p^{2}}{4m}\), where  m is the paticle mass.  In addition,
we assume that the pair is created with zero total angular momentum.
Hence, the wave function can be written in the following form:

\begin{equation}
  \psi=\sum_{nlm}\frac{Y_{lm}(\vec{r})Y_{lm}^{*}(\vec{\varrho})}{\sqrt{2l+1}}
\psi_{nl}
\end{equation}
and the radial  wave functions are assumed in the  form
\begin{equation}
  \psi_{nl}=\varphi_{nl}(r)\phi_{nl}(\varrho)
\end{equation}
\begin{equation}
  \phi_{nl}= j_{l}(p\varrho) a_{nl}(\varrho)
\end{equation}
where the unknown functions $a_{nl}(\varrho)$ are to be found .
The form of this expansion  although general has been chosen for the
high-energy
region which allows simple approximate solutions for the functions $a$ .
After eliminating the internal degrees of freedom by integration
over  r one  obtains  an infinite set of  second order linear differential
equations for $a_{nl}$
that involve effective potentials in the "external" variable $\varrho$

\begin{equation}
  V_{nln'l'}(\varrho)=\sum_{\lambda=1,3...}\frac{<\varphi_{nl}\mid V_{\lambda}
  \mid\varphi_{n'l'}>}{\sqrt{(2l+1)(2l'+1)}} Delta(l,l',\lambda)
\end{equation}

where
\begin{equation}
Delta(l,l',\lambda)=(2l+1)(2l'+1)(2\lambda+1) \left(\begin{array}{ccc}
l & l' &\lambda \\ 0 & 0 & 0 \end{array} \right)^{2}
\end{equation}
and
\(\left(\begin{array}{ccc}
l & l' &\lambda \\ 0 & 0 & 0 \end{array} \right)^{2}\) is the Wigner symbol.

These  potentials are related to the  external Coulomb potential V  by the
multipole expansion :

\begin{equation}
  V(\vec{r},\vec{\varrho})=\sum_{\lambda m} V_{\lambda}Y_{\lambda m}
  ^{*}(\vec{r})Y_{\lambda m}(\vec{\varrho})
\end{equation}
with  $V_{\lambda}$ the radial part of the ${\lambda}$-pole
polarisation potential given by
\begin{equation}
V_{\lambda}(r,\varrho)=\frac{4M \alpha}{2\lambda+1}
\frac{x_{<}^{\lambda}}{x_{>}^{\lambda+1}}, \  \lambda=1,3...
\end{equation}
where x is $\varrho$ or $\frac{r}{2}$ .

   The dominant dipole effective potentials are long-ranged
of the $\varrho^{-2}$ asymptotics . At distances close to the
atomic radii these potentials fall down sharply and exhibit zeros
of high order  at the origin . Few examples are given in  Appendix A.
The long range of these potentials allows some vital approximations in the
high energy region . We refer to Appendix B where the details are
discussed in the language of integral equations while some numerical
studies are performed in the next section .One effect of the range
is some  smoothness of the functions  $a$ and this allows for
differential equations of the first order. Another special feature
is related both to the long-range and the angular momentum
carried by the effective potential . This allows for the expansion (6,7)
where , one  would otherwise expect also some hankel functions corresponding
to ingoing( outgoing ) waves . The first order equation discussed in
Appendix B is

\begin{equation}
  a_{nl}'=\sum_{n'l'}\frac{Z}{2p} V_{nln'l'}a_{n'l'}(-1)^{l'}
\end{equation}
where the length is expressed in the Bohr radius units .
These  equations for $a_{nl}(\varrho)$ can be written in the following
matrix form
\begin{equation}
a'=Ua
\end{equation}
or
\begin{equation}
ia'=iUa
\end{equation}
where $U$ is an antisymmetric matrix and respectively $iU$ is hermitian.
The boundary conditions can be formulated via integral equation
\begin{equation}
a=a_{0}+\int_{\varrho}^{\infty}Ua \ d\varrho
\end{equation}
which is the set of integral equations derived in Appendix B .
For  example for the incoming wave coresponding with the state 1S
\begin{equation}
a_{0}=(1,0,0,...)
\end{equation}
and the vector a represents following descrete states
\begin{equation}
a=(1S,2S,2P,3S,3P,3D,...)
\end{equation}

\section{Illustrative example -  two discrete states }

  To illustrate the  formalism introduced above let us consider an
idealised atomic system of only two:  2S and 2P states .
It is easy to find ,in this case , the exact as well as the approximate
solutions for the integral or differential equations for the  final
atomic state of the pair . If the incident wave is formed with
the 2S state one obtains from eq(12) or eq(15)  :

\begin{eqnarray}
a_{20}(\varrho)=cos(\frac{Z}{2 p} \int_{\varrho}^{\infty} V_{2120})
\end{eqnarray}
which generates the  S wave amplitude  of the pair  with respect to the
formation region  . Now the 2S atom formation probablity pr(2S) is given by
the square of the amplitude at the origin
\begin{eqnarray}
pr(2S)=A cos^{2}(\frac{Z}{2 p} \int_{\varrho}^{\infty} V_{2120})
\end{eqnarray}
where A is a constant involving pair production amplitude and
some phase space integrations , [ see 6 for more details].

In the same way one can solve  for an amplitude $a_{20}$ in the case
of the incident 2P state
\begin{eqnarray}
a_{20}(\varrho)=sin(\frac{Z}{2 p} \int_{\varrho}^{\infty} V_{2120})
\end{eqnarray}
and the probability of the 2P atomic state formation is
\begin{eqnarray}
pr(2P)=A sin^2(\frac{Z}{2 p} \int_{\varrho}^{\infty} V_{2120})
\end{eqnarray}
The total probability of the atom formation is then energy independent
\begin{eqnarray}
pr=pr(2S)+pr(2P)=A
\end{eqnarray}
and  equals in the high energy limit to the 2S state formation
probability given by the  coalescence model .
In figs 1,2  parts of the plots of pr(2S) and pr(2P) are presented for a
special case
of pozitonium formed in the strong field ( Z=170) of heavy ions .

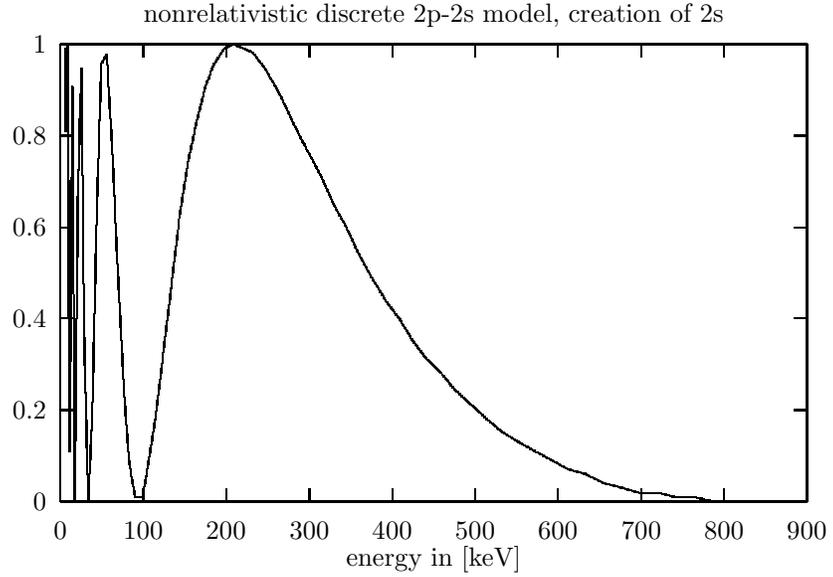
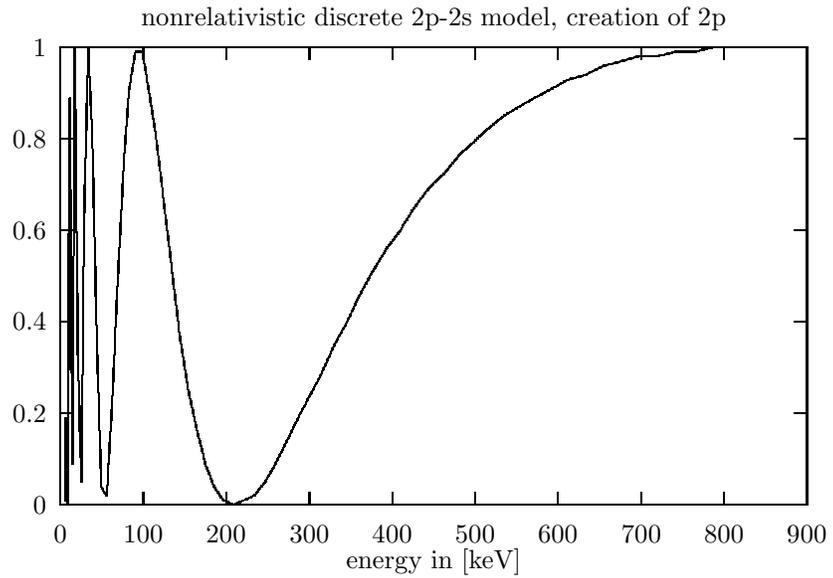
\input{f1}

\input{f2}

Let us notice several essential points.

 1) The oscillations of the atomic production probability reflect
 quantum beating of this system . In the strong external electric field
 there are two diagonal combinations  S(i)S(e)+P(i)P(e)   and
 S(i)S(e)-P(i)P(e) where (i) denotes the internal atomic state and (e) the
 state of the external angular momentum of the atom . These two components
 beat in time  and  in space when the atom propagates through the
 polarising electric field.

 2) This simple case may be solved exactly . We find that our high energy
 approximation works well for kinetic energies higher than the
 barrier inherent in the effective potential . In practice it covers
 the region of three last peaks in the formation probability .

 3) The beating effects persist at kinetic energies higher than the atomic
 binding energies by orders of magnitude .  The Coulomb potential and
 the dominant dipole polarising potential have no inherent
 scale parameter. This parameter is introduced via the atomic matrix
 elements (8) that determine the effective V . Thus the energy scale in
 fig 1 is determined by the dimensionless parameter Z/pB .
 It is much larger than the scale of the atomic binding energies .

   The question to be answered is to what extent this beating of the
Stark mixed  atomic state persists in  complete calculations .
Such calculations should allow decays of the incident state to higher
states and to the continuum . We study it by expanding the
basis of the discrete levels allowed and by an approximate treatment
of the continuum .The latter is discussed in the next section  .

\section{Approximate treatment of continuum }

In appendix B we estimate the continuum part $ \psi^{c} $
of the wave function $\psi$ . This part is related to the continuous
spectrum involved in the  Coulomb Green function  $ G^{c} $  entering
the Lippman-Shwinger eq(3) .
We argue that at high energies there exist an approximate relation
\begin{equation}
<\varphi_{nlm}\mid V G^{c}V\mid \varphi_{n'l'm'}> \  \approx \
<\varphi_{nlm}\mid V \delta (\vec{r}-\vec{r'})G_{p}V\mid \varphi_{n'l'm'}>
\end{equation}
Then we expand the delta function into a system of eigenstates
\begin{equation}
\delta(\vec{r}-\vec{r'})=\sum_{nlm} \mid \psi_{nlm}><\psi_{nlm} \mid+
continuous \ \ part
\end{equation}
and  neglect the continuous terms to obtain the following approximation
\begin{equation}
<\varphi_{nlm}\mid V G^{c}V\mid \varphi_{n'l'm'}> \  \approx \
<\varphi_{nlm}\mid V G^{d}V\mid \varphi_{n'l'm'}>
\end{equation}
where $ G^{d} $ is the discrete part of Coulomb Green's function.
In this way , we expand the continuous piece of the  wave function
$\psi^{c}$
into discrete states in the same way as it was  done in eqs (5,6,7)
with the $\psi^{d}$ . The coefficients of this expansion are named $ b $ .
Now defining a new vector $\tilde{a}$
\begin{equation}
\tilde{a}=(a,b)
\end{equation}
we obtain the equation
\begin{equation}
\tilde{a}'=\tilde{U}\tilde{a}
\end{equation}
where the new potential $ \tilde{U} $ is defined as follows:
\begin{equation}
\tilde{U}=
\left(\begin{array}{cc}
U & U \\
U & 0
   \end{array} \right)
\end{equation}
The equation for $ \tilde{a} $  is soluble  with simple
Runge-Kutta methods. This procedure doubles the dimension of
the previously described
set of differential equations .Its effect on the discrete part
$\psi^{d} $ is equivalent to an optical potential contribution
leading to the equation
\begin{equation}
\psi^{d}=\psi^{d}_{0}+G^{d}V\psi^{d}+G^{d}VG^{d}V\psi^{d}
\end{equation}
It turns out that contributions from $\psi^{c}$ at the origin vahish
quickly with increasing energy and for intermediate energies stop plaing
an important role.

\section{Calculations and Conclusions}

This calculation  decribes  formation of atomic states  of  a pair
$e^{+} \ e^{-}$ , $\pi^{+} \ \pi^{-}$ etc. in the electric field of
colliding nuclei that create the pair. It is shown that the rate of
creation of pairs in a given atomic state depends strongly on the final
energy of the pair. This result depends on one  scale
parameter $\frac{2 M Z \alpha}{2 p}$
so it can be rescaled for different masses of the pair, total atomic
numbers of colliding nuclei $Z$ and  energies. The  probability to form
the  1S state changes rather smoothly  with energy from negligible values at
low energies to the coalescence model result at high energies .
 A reliable calculation of the formation rates  in higher atomic states
 requires  a large number of atomic states in the basis of expansion .
 We have solved the system of differential equations  up to K= 21 states.
 Good convergence of probabilities with respect to K have been obtained
 only for the first three states.
 Within higher shells some  traces of the beating effects are still
 visible  although these are  rather unstable with respect to K .
 This question requires further studies and possibly better approximations
 as for large K the effective potentials create numerical difficulties .

\bigskip
\noindent {\bf Acknowledgements}

This work was supported in part by the KBN grant 2P302 140 04

{\bf References}

1. H Nann in Proc . Workshop on Meson Production ,interaction and decay.
 Cracow Poland 1991, Ed.Magiera A,W.Oelert,E.Grosse,World Scient. 91,p100

2. W.Oelert , ibid p 199

3  R.Siebert , ibid p 166

4. Bern-Dubna Collab ,CERN Letter of Intent SPSLC92-44/I191

5. S.Vigdor , Indiana U C L Proposal 1993

6. S.Wycech , AM. Green , Nucl Phys A562(1993) 446

7. O.Dumbrajs, Zeit.Phys A321(1985)297

8. J. Schweppe, A Ruppe et al.  PRL {\bf 51}  (1983)  2261,  PRL  {\bf 54}
(1985) 1761

\section{Appendix A - Wave functions and potentials}
The first three wave functions and effective potentials are given below .
The atomic units ( B =1) are used.The radial  wave functions :

\begin{equation}
\varphi_{10}= {2 {{e^{-r}}}}
\end{equation}

\begin{equation}
\varphi_{20}= {{(2 - r)}  \over {2 {\sqrt{2}}}} e^{-{r\over 2}}
\end{equation}

\begin{equation}
\varphi_{21}= {r\over {2 {\sqrt{6}}}} e^{-{r\over 2}}
\end{equation}
generate , under the dipole external field , the following
transition potentials :

\begin{equation}
V_{2110}(x) = \frac{(5.959 + (-5.959 - 17.88 x - 26.82 x^2 - 20.11 x^3)
        e^{-3 x}}{x^2}
\end{equation}
\begin{equation}
V_{2120}(x) = \frac{(-24. + (24. + 48. x + 48. x^2 + 32. x^3 + 16. x^4)
        e^{-2 x}}{x^2}
  \end{equation}
\begin{equation}
V_{3021}(x) = \frac{(4.33 + (-4.33 - 7.22 x - 6.02 x^2 - 3.17 x^3 -
    1.11 x^4 - 3.71 x^5) e^{-5 x}}{ 3 x^2}
 \end{equation}

All other transition potentials  defined in section 1
can be found in explicit form as elementary functions.

\section{Appendix B - High energy approximations }
The  wave function $\psi$ can be divided into two parts -  discrete
and  continuous:
\begin{equation}
\psi = \psi^{d} +\psi^{c}
\end{equation}
This separation follows the same seperation of the Coulomb
Green function.
\begin{equation}
G=G^{d}+G^{c}
\end{equation}
Where the first part contains the atomic  and the second part the
continuous internal states of the pair .
Now Lippman-Schwinger equation is equivalent to coupled equations for
the two parts of of $\psi$.
\begin{equation}
\begin{array}{c}
\psi^{d}=\psi^{d}_{0}+G^{d}V\psi^{d}+G^{d}V\psi^{c} \\

\psi^{c}= G^{c}V\psi^{d}+G^{c}V\psi^{c}
\end{array}
\end{equation}
In this calculation  we are interested in atomic states , the
continuum  will be described in a simplified way similar to
 the  optical potential method . Thus we neglect
the  $G^{c}V\psi^{c}$ term that changes the continuum part of the
propagator . Now
\begin{equation}
\psi^{c}=G^{c}V\psi^{d}
\end{equation}
is substituted into equation for $\psi^{c}$ that generates the optical
model equation
\begin{equation}
\psi^{d}=\psi^{d}_{0}+G^{d}V\psi^{d}+G^{d}VG^{c}V\psi^{d}
\end{equation}
For high energies, much larger than the atomic binding energies ,
one has an approximate relation for the matrix element
\begin{equation}
<\varphi_{nlm}\mid V G^{c}V\mid \varphi_{n'l'm'}> \  \approx \
<\varphi_{nlm}\mid V G_{0}V\mid \varphi_{n'l'm'}>
\end{equation}
so in the above equations we  interchange $G^{c}$ with $G_{0}$.
The discrete part of propagator has the following form
\begin{equation}
    G^{d}= G^{d}(\vec{r},\vec{\varrho},\vec{r'},\vec{\varrho'}) =
    \sum_{nlm} \mid\varphi_{nlm}>< \varphi_{nlm}\mid G_{p}(\vec{\varrho},
    \vec{\varrho'},E-E_{n}),
\end{equation}
where for high energies E  we neglect the binding energies $E_{n}$.
In a similar fashion for small relative momenta of the pair
the Green function $G_{0}$ will be used in the following approximate
form
\begin{equation}
 G_{0}(\vec{r},\vec{\varrho},\vec{r'},\vec{\varrho'}) =
   \delta(\vec{r}-\vec{r'}) G_{p}(\vec{\varrho}, \vec{\varrho'}),
 \end{equation}

   The external potential changes the internal angular momentum . This
is crucial for our problem and allows an  approximation that we discuss
now . Let us go over to the angular momentum representation and follow
the expansion given in eqs(5) and (6) .
The propagator is also expanded into spherical harmonics
\begin{equation}
 G_{p}(\vec{\varrho},\vec{\varrho'})= \sum_{lm}g_{l}(\varrho,\varrho')
 Y^{*}_{lm}(\vec{\varrho}) Y_{lm}(\vec{\varrho'})
 \end{equation}
with
\begin{equation}
g_{l}(\varrho,\varrho')= - i p j_{l}(p\varrho_{<})h_{l}(\varrho_{>})
\end{equation}
Now we can obtain the system of
integral equations for the radial functions $\phi $ of eg(6)
\begin{eqnarray}
\phi_{nl}(\varrho)=j_{l_{0}}(p \varrho) \delta_{nn_{0}} \delta_{ll_{0}}+
Z \int g_{l}(\varrho,\varrho')
\sum_{n'l'}V_{nln'l'} \phi_{nl}(\varrho')  \nonumber \\
+ Z^2 \int g_{l}(\varrho,\varrho')
\sum_{n'l' \alpha}W_{nln'l'}^{\alpha}(\varrho',\varrho'')
g_{\alpha}(\varrho',\varrho'')
\phi_{n'l'}(\varrho'')
\end{eqnarray}
The transition potential $V_{nln'l'}$ is defined in the following way:

\begin{equation}
  V_{nln'l'}(\varrho)=\sum_{\lambda=1,3...}\frac{<\varphi_{nl}\mid V_{\lambda}
  \mid\varphi_{n'l'}>}{\sqrt{(2l+1)(2l'+1)}} Delta(l,l',\lambda)
\end{equation}
and $W_{nln'l'}^{\alpha}$ is given by the formula
\begin{eqnarray}
  W_{nln'l'}^{\alpha}(\varrho,\varrho')= \ \ \ \ \ \ \ \ \  \ \nonumber \\
  \sum_{\lambda,\lambda'=1,3...}
  \frac{<\varphi_{nl}\mid V_{\lambda}(\varrho,r) V_{\lambda'}(\varrho',r)
\mid\varphi_{n'l'}>}{(2\alpha+1)\sqrt{(2l+1)(2l'+1)}} Delta(l,\alpha,\lambda)
  Delta(l',\alpha,\lambda')
\end{eqnarray}

  Iterations of these integral equations involve products of the bessel
 and neuman functions . These are integrated with the effective
 potentials  V which are predominantly long-ranged .
 For high energies  i.e. for high momenta p these functions oscillate
 rapidly and one obtains approximate relations for mean values:
\begin{equation}
< j_{l}(x)j_{l'}(x)> \approx \frac{1}{2x^{2}}cos((l-l')\frac{\pi}{2})
\end{equation}
\begin{equation}
< n_{l}(x)n_{l'}(x)> \approx \frac{1}{2x^{2}}cos((l-l')\frac{\pi}{2})
\end{equation}
\begin{equation}
< j_{l}(x)n_{l'}(x)> \approx \frac{1}{2x^{2}}sin((l-l')\frac{\pi}{2})
\end{equation}
where
\begin{equation}
x=p\varrho
\end{equation}
and symbols $<f(x)>$ denote the mean value of an oscillating function f(x)
weihted by V.
We repeat again , that the success of these approximation is related to
the long range of the dipole polarising potentials V .
When we write $\phi_{nl}$  in the standard form
\begin{equation}
  \phi_{nl}= j_{l}(p\varrho) a_{nl}(\varrho)+h_{l}(p\varrho) b_{nl}(\varrho)
\end{equation}
and  apply the expressions for mean values of bessel functions we obtain
\begin{eqnarray}
a_{nl}(\varrho)= \delta_{nn_{0}} \delta_{ll_{0}}+
\int_{\varrho}^{\infty} \frac{1}{2p}sin((l-l')\frac{\pi}{2})
\sum_{n'l'}V_{nln'l'} \a_{nl}(\varrho')d \varrho'  \nonumber \\
+ \int_{\varrho}^{\infty} \frac{1}{4p^{2}}sin((\alpha-l)\frac{\pi}{2})
sin((l'-\alpha)\frac{\pi}{2})
\sum_{n'l' \alpha}W_{nln'l'}^{\alpha}(\varrho',\varrho'')
\a_{n'l'}(\varrho'')d\varrho' d\varrho''
\end{eqnarray}
The equation for $b_{nl}$ has the similar form:
\begin{eqnarray}
b_{nl}(\varrho)=
\int_{0}^{\varrho} \frac{1}{2p}sin((l-l')\frac{\pi}{2})
\sum_{n'l'}V_{nln'l'} \a_{nl}(\varrho')d \varrho'  \nonumber \\
+ \int_{0}^{\varrho} \frac{1}{4p^{2}}sin((\alpha-l)\frac{\pi}{2})
sin((l'-\alpha)\frac{\pi}{2})
\sum_{n'l' \alpha}W_{nln'l'}^{\alpha}(\varrho',\varrho'')
b_{n'l'}(\varrho'')d\varrho' d\varrho''
\end{eqnarray}
but because there is no initial term $ \delta_{nn_{0}} \delta_{ll{0}}$ the
solution of
the above equation is trivial - i.e. $b_{nl}=0$.
The equation for $a_{nl}$ is the gateway for the high energy approximation.

\end{document}